\documentclass[aps,prd,twocolumn,superscriptaddress,showpacs]{revtex4}

\usepackage{epsfig}
\usepackage{psfrag}
\usepackage{graphicx}% Include figure files
\usepackage{dcolumn}% Align table columns on decimal point
\usepackage{bm}% bold math
\usepackage{epsfig}
\usepackage{multirow}
\usepackage{slashed}
\usepackage{rotating}
\usepackage{subfig}
\usepackage{caption}
\usepackage{verbatim} 
\usepackage{mathrsfs} 

\newcommand{\pt}{\mbox{$p_{T}$}}
\newcommand{\mjl}{\mbox{$D^{M_{\text{SV}}}_{\text{JLIP}}$}}

\newcommand{\mll}{\mbox{$m_{\ell \ell}$}}

\newcommand{\dzero}{D0}

\newcommand{\ee}      {\ensuremath{ee}}
\newcommand{\mumu}    {\ensuremath{\mu\mu}}

\newcommand{\GeV} {\ensuremath{\mathrm{Ge\kern -0.1em V}}}
\newcommand{\TeV} {\ensuremath{\mathrm{Te\kern -0.1em V}}}

\newcommand{\ppbar}{\mbox{$p\overline{p}$}}

\newcommand{\ttbar}{\mbox{$t\overline{t}$}}

\newcommand{\alpgen}{{\sc alpgen}}
\newcommand{\pythia}{{\sc pythia}}
\newcommand{\mcfm}{{\sc mcfm}}
\newcommand{\geant}{{\sc geant3}}

\begin{document}
%\linenumbers

\hspace{5.2in} \mbox{FERMILAB-PUB-10-435-E}

\title{\boldmath 
A measurement of the ratio of inclusive cross sections $\sigma(p\bar{{p}}\rightarrow Z+b{\rm\, jet})/
\sigma(p\bar{{p}}\rightarrow Z+{\rm jet})$ at $\sqrt{{s}}=1.96$ TeV}

\affiliation{Universidad de Buenos Aires, Buenos Aires, Argentina}
\affiliation{LAFEX, Centro Brasileiro de Pesquisas F{\'\i}sicas, Rio de Janeiro, Brazil}
\affiliation{Universidade do Estado do Rio de Janeiro, Rio de Janeiro, Brazil}
\affiliation{Universidade Federal do ABC, Santo Andr\'e, Brazil}
\affiliation{Instituto de F\'{\i}sica Te\'orica, Universidade Estadual Paulista, S\~ao Paulo, Brazil}
\affiliation{Simon Fraser University, Vancouver, British Columbia, and York University, Toronto, Ontario, Canada}
\affiliation{University of Science and Technology of China, Hefei, People's Republic of China}
\affiliation{Universidad de los Andes, Bogot\'{a}, Colombia}
\affiliation{Charles University, Faculty of Mathematics and Physics, Center for Particle Physics, Prague, Czech Republic}
\affiliation{Czech Technical University in Prague, Prague, Czech Republic}
\affiliation{Center for Particle Physics, Institute of Physics, Academy of Sciences of the Czech Republic, Prague, Czech Republic}
\affiliation{Universidad San Francisco de Quito, Quito, Ecuador}
\affiliation{LPC, Universit\'e Blaise Pascal, CNRS/IN2P3, Clermont, France}
\affiliation{LPSC, Universit\'e Joseph Fourier Grenoble 1, CNRS/IN2P3, Institut National Polytechnique de Grenoble, Grenoble, France}
\affiliation{CPPM, Aix-Marseille Universit\'e, CNRS/IN2P3, Marseille, France}
\affiliation{LAL, Universit\'e Paris-Sud, CNRS/IN2P3, Orsay, France}
\affiliation{LPNHE, Universit\'es Paris VI and VII, CNRS/IN2P3, Paris, France}
\affiliation{CEA, Irfu, SPP, Saclay, France}
\affiliation{IPHC, Universit\'e de Strasbourg, CNRS/IN2P3, Strasbourg, France}
\affiliation{IPNL, Universit\'e Lyon 1, CNRS/IN2P3, Villeurbanne, France and Universit\'e de Lyon, Lyon, France}
\affiliation{III. Physikalisches Institut A, RWTH Aachen University, Aachen, Germany}
\affiliation{Physikalisches Institut, Universit{\"a}t Freiburg, Freiburg, Germany}
\affiliation{II. Physikalisches Institut, Georg-August-Universit{\"a}t G\"ottingen, G\"ottingen, Germany}
\affiliation{Institut f{\"u}r Physik, Universit{\"a}t Mainz, Mainz, Germany}
\affiliation{Ludwig-Maximilians-Universit{\"a}t M{\"u}nchen, M{\"u}nchen, Germany}
\affiliation{Fachbereich Physik, Bergische  Universit{\"a}t Wuppertal, Wuppertal, Germany}
\affiliation{Panjab University, Chandigarh, India}
\affiliation{Delhi University, Delhi, India}
\affiliation{Tata Institute of Fundamental Research, Mumbai, India}
\affiliation{University College Dublin, Dublin, Ireland}
\affiliation{Korea Detector Laboratory, Korea University, Seoul, Korea}
\affiliation{CINVESTAV, Mexico City, Mexico}
\affiliation{FOM-Institute NIKHEF and University of Amsterdam/NIKHEF, Amsterdam, The Netherlands}
\affiliation{Radboud University Nijmegen/NIKHEF, Nijmegen, The Netherlands}
\affiliation{Joint Institute for Nuclear Research, Dubna, Russia}
\affiliation{Institute for Theoretical and Experimental Physics, Moscow, Russia}
\affiliation{Moscow State University, Moscow, Russia}
\affiliation{Institute for High Energy Physics, Protvino, Russia}
\affiliation{Petersburg Nuclear Physics Institute, St. Petersburg, Russia}
\affiliation{Stockholm University, Stockholm and Uppsala University, Uppsala, Sweden }
\affiliation{Lancaster University, Lancaster LA1 4YB, United Kingdom}
\affiliation{Imperial College London, London SW7 2AZ, United Kingdom}
\affiliation{The University of Manchester, Manchester M13 9PL, United Kingdom}
\affiliation{University of Arizona, Tucson, Arizona 85721, USA}
\affiliation{University of California Riverside, Riverside, California 92521, USA}
\affiliation{Florida State University, Tallahassee, Florida 32306, USA}
\affiliation{Fermi National Accelerator Laboratory, Batavia, Illinois 60510, USA}
\affiliation{University of Illinois at Chicago, Chicago, Illinois 60607, USA}
\affiliation{Northern Illinois University, DeKalb, Illinois 60115, USA}
\affiliation{Northwestern University, Evanston, Illinois 60208, USA}
\affiliation{Indiana University, Bloomington, Indiana 47405, USA}
\affiliation{Purdue University Calumet, Hammond, Indiana 46323, USA}
\affiliation{University of Notre Dame, Notre Dame, Indiana 46556, USA}
\affiliation{Iowa State University, Ames, Iowa 50011, USA}
\affiliation{University of Kansas, Lawrence, Kansas 66045, USA}
\affiliation{Kansas State University, Manhattan, Kansas 66506, USA}
\affiliation{Louisiana Tech University, Ruston, Louisiana 71272, USA}
\affiliation{Boston University, Boston, Massachusetts 02215, USA}
\affiliation{Northeastern University, Boston, Massachusetts 02115, USA}
\affiliation{University of Michigan, Ann Arbor, Michigan 48109, USA}
\affiliation{Michigan State University, East Lansing, Michigan 48824, USA}
\affiliation{University of Mississippi, University, Mississippi 38677, USA}
\affiliation{University of Nebraska, Lincoln, Nebraska 68588, USA}
\affiliation{Rutgers University, Piscataway, New Jersey 08855, USA}
\affiliation{Princeton University, Princeton, New Jersey 08544, USA}
\affiliation{State University of New York, Buffalo, New York 14260, USA}
\affiliation{Columbia University, New York, New York 10027, USA}
\affiliation{University of Rochester, Rochester, New York 14627, USA}
\affiliation{State University of New York, Stony Brook, New York 11794, USA}
\affiliation{Brookhaven National Laboratory, Upton, New York 11973, USA}
\affiliation{Langston University, Langston, Oklahoma 73050, USA}
\affiliation{University of Oklahoma, Norman, Oklahoma 73019, USA}
\affiliation{Oklahoma State University, Stillwater, Oklahoma 74078, USA}
\affiliation{Brown University, Providence, Rhode Island 02912, USA}
\affiliation{University of Texas, Arlington, Texas 76019, USA}
\affiliation{Southern Methodist University, Dallas, Texas 75275, USA}
\affiliation{Rice University, Houston, Texas 77005, USA}
\affiliation{University of Virginia, Charlottesville, Virginia 22901, USA}
\affiliation{University of Washington, Seattle, Washington 98195, USA}
\author{V.M.~Abazov} \affiliation{Joint Institute for Nuclear Research, Dubna, Russia}
\author{B.~Abbott} \affiliation{University of Oklahoma, Norman, Oklahoma 73019, USA}
\author{B.S.~Acharya} \affiliation{Tata Institute of Fundamental Research, Mumbai, India}
\author{M.~Adams} \affiliation{University of Illinois at Chicago, Chicago, Illinois 60607, USA}
\author{T.~Adams} \affiliation{Florida State University, Tallahassee, Florida 32306, USA}
\author{G.D.~Alexeev} \affiliation{Joint Institute for Nuclear Research, Dubna, Russia}
\author{G.~Alkhazov} \affiliation{Petersburg Nuclear Physics Institute, St. Petersburg, Russia}
\author{A.~Alton$^{a}$} \affiliation{University of Michigan, Ann Arbor, Michigan 48109, USA}
\author{G.~Alverson} \affiliation{Northeastern University, Boston, Massachusetts 02115, USA}
\author{G.A.~Alves} \affiliation{LAFEX, Centro Brasileiro de Pesquisas F{\'\i}sicas, Rio de Janeiro, Brazil}
% ...
%
% visitor_addresses.tex                        6 October 2010
%  available symbols are:
%  $\ast, \dag, \ddag, \S, \P, $\|$, $\ast\ast$, \dag\dag, \ddag\ddag ,\#
%
\collaboration{The D0 Collaboration\footnote{with visitors from
%{alton}
$^{a}$Augustana College, Sioux Falls, SD, USA,
%{burdin}
$^{b}$The University of Liverpool, Liverpool, UK,
%{haas,partridge}
$^{c}$SLAC, Menlo Park, CA, USA,
%{juste}
$^{d}$ICREA/IFAE, Barcelona, Spain,
%{luna-garcia}
$^{e}$Centro de Investigacion en Computacion - IPN, Mexico City, Mexico,
%{podesta-lerma}
$^{f}$ECFM, Universidad Autonoma de Sinaloa, Culiac\'an, Mexico,
and 
%{weber}
$^{g}$Universit{\"a}t Bern, Bern, Switzerland.%
%{hooper}
%$^{?}$%Visitor from Bradley University, Peoria, IL, USA.
%{kozminski
%$^{?}$}%Visitor from Lewis University, Romeoville, IL, USA.
%{deceased}
%$^{\ddag}$%Deceased.
}} \noaffiliation
\vskip 0.25cm

\date{October 29, 2010}

\begin{abstract}
% remove the space for publication
\vspace*{3.0cm} The ratio of the cross section for $p\bar{p}$ interactions producing a 
$Z$ boson and at least one $b$ quark jet
to the inclusive $Z+{\rm jet}$ 
cross section is measured using $4.2\ {\rm fb}^{-1}$ of $p\bar{p}$ collisions
collected with the \dzero\ detector at the Fermilab Tevatron
collider at $\sqrt{{s}}=1.96$ TeV.
The $Z\rightarrow\ell^+\ell^-$ 
candidate events with
at least one $b$ jet are discriminated from $Z+$~charm and light jet(s)
events by a novel technique that exploits the properties of the tracks associated to the jet. The measured ratio
is $0.0193\pm0.0027$
for events having a jet with transverse momentum $\pt > 20~\GeV$
and pseudorapidity $|\eta| \leq 2.5$, which is the most precise 
to date and is consistent with theoretical predictions.

% remove this for publication
\end{abstract}
\pacs{12.38.Qk, 13.85.Qk, 14.65.Fy, 14.70.Hp}
\maketitle

The measurement of the production cross section for a $Z$ boson in association
with $b$ jets provides an important test of perturbative quantum chromodynamics (QCD)
predictions~\cite{Campbell}. A good description of this process by theoretical calculations is essential
since it is a major background to searches for the standard model (SM) Higgs boson via 
$ZH(H\rightarrow b\bar{b})$ associated production~\cite{zhllbb} and for the supersymmetric
partners of $b$ quarks~\cite{sbottom}. This process is also sensitive to
the $b$ quark density in the proton needed to predict
phenomena such as single top quark production~\cite{singleTop} and production of
non-SM Higgs bosons in association with $b$ quarks~\cite{susyHiggs}.
Calculations for the $Z$ boson production in association 
with $b$ quarks in $p\bar{p}$ collisions are available at 
next-to-leading order (NLO) using two different 
approaches~\cite{Campbell,Doreen}, and they agree 
within their respective theoretical uncertainties.
%\newpage

In this Letter, we describe a measurement of the ratio of the 
inclusive cross sections for $Z$ boson production with at least one $b$ quark jet 
to the $Z+{\rm jet(s)}$ production in $p\bar{p}$ interactions, 
where the $Z$ boson is identified via its $Z \rightarrow ee$ and $Z \rightarrow \mu\mu$ decay modes.
The $Z+b{\rm\, jet}$ events are separated from $Z$ boson production with light 
($u$, $d$, or $s$ quarks, or gluons) and charm ($c$) jet(s)
by a discriminant that exploits the properties of the tracks associated to the jet.
The measurement of the ratio benefits from cancellations of many systematic 
uncertainties on the cross sections and therefore allows a more precise comparison with theoretical calculations.
Previous measurements by the \dzero~\cite{D0Paper} and CDF~\cite{CDFPaper} collaborations
agree with the SM predictions. 
Here, we present the most precise measurement of the ratio to date.
This measurement is a significant improvement over the previous
\dzero\ result~\cite{D0Paper}  which utilized 0.18~fb$^{-1}$ of integrated luminosity and  
assumed the ratio of the $Z+b$ jet cross section to the $Z+c$ jet cross section from NLO calculations.
This analysis uses a much larger dataset and a substantially improved method 
to extract the different jet flavor fractions.

We use data from $p\bar{p}$ collisions at a center-of-mass 
energy of 1.96 TeV collected by the \dzero\ detector~\cite{d0det} at the Fermilab Tevatron 
between 2006 and 2009 and corresponding to an integrated luminosity 
of 4.2 fb$^{-1}$. The selected
events are required to pass at least one of the single electron or single muon triggers.
The efficiency of the triggers, as  measured from data,
is close to 100\% (78\%) for the $Z \rightarrow ee$ ($Z \rightarrow \mu\mu$) final state.

This analysis relies on all components of the detector: tracking, 
calorimetry, and muon system and the ability to identify detached vertices.
The \dzero\ detector consists of a central tracking system,
comprising a silicon microstrip tracker (SMT) and a central
fiber tracker, both within a 2~T solenoidal magnet; a
liquid-argon and uranium calorimeter, divided into a central
calorimeter and two endcap calorimeters; and a muon
system, consisting of three layers of tracking detectors
and scintillation trigger counters. 
The SMT allows a precise reconstruction of the $p\bar{p}$ interaction vertex (PV)
and of eventual secondary vertices (\rm{SV}), and 
an accurate determination of the impact parameter (\rm{IP}) of a track relative to
the PV, which are the key components of the jet lifetime-based $b$-tagging algorithms.

Offline event selection requires a reconstructed PV that has
at least three associated tracks and is located within 60 cm of the center
of the detector in the coordinate along the beam direction. The selected events 
must contain a $Z$ boson candidate with a dilepton invariant mass $70~\GeV < m_{\ell \ell}
< 110~\GeV$. Throughout this Letter we use $Z$ boson to denote any dilepton event in the
above mentioned mass range due to $Z$ or $\gamma^*$ production.

The dielectron ($ee$) selection requires at least two electrons of 
transverse momentum $\pt > 15~\GeV$ identified by electromagnetic (EM) showers
in the central (with pseudorapidity~\cite{def} $|\eta|<1.1$) or 
endcap ($1.5<|\eta|<2.5$) calorimeter. The showers must have a significant
fraction of their energy deposited in the EM calorimeter, be isolated from 
other energy depositions, and have a shape consistent with that expected for an electron.
The central electrons, in addition, must
match central tracks or produce electron-like patterns of hits in the
tracker.

The dimuon ($\mu\mu$) selection requires at least two muons 
with segments in the muon spectrometer matched to central tracks with $\pt > 10~\GeV$ and $|\eta|<2$.
Combined tracking and calorimeter isolation requirements are applied to the muon candidates.
Muons from cosmic rays are rejected by applying a timing criterion to the hits in the
scintillator layers as well as restricting the position of the muon track with respect to the PV.
The two muons must also have opposite electric charges.

A total of 411,064 (224,814) $Z$ boson candidate events are retained in the $ee$ ($\mu\mu$) channel.
The $Z+{\rm jet}$ sample is then selected by requiring the presence of at least 
one reconstructed jet with $|\eta|<2.5$, with the leading jet having 
$\pt > 20~\GeV$ and any additional jets having $\pt > 15~\GeV$.
Jets are reconstructed from energy deposits in the calorimeter using the iterative midpoint cone
algorithm \cite{RunIIcone} with a cone of radius 0.5.
The energy of
jets is corrected for detector response, the presence of noise and multiple
$p\bar{p}$ interactions, and the energy deposited outside of the jet cone used for reconstruction.
Jets containing $b$ quarks have a different energy response and receive
an average additional energy correction of about 6\% as determined 
from simulations. Events with missing transverse
energy larger than 60~\GeV\ 
are rejected to suppress the background from $\ttbar$ 
production.
These selection criteria yield a sample of 48,956 (24,450) $Z+$jet events
in the $ee$ ($\mu\mu$) channel.

Jets considered for $b$-tagging are subject to a preselection, called taggability, 
to decouple the intrinsic $b$ jet tagging algorithm performance from other effects.
For this purpose, the jet is required to have at least two associated
tracks with $\pt > 0.5~\GeV$, the leading track must 
have  $\pt > 1.0~\GeV$, and each track
must have at least one SMT hit. 
This requirement has a typical efficiency of 90\% per jet. 
The jet related efficiencies mentioned here and later on are determined from
simulations and corrected for the difference observed in data.
In order to enrich a sample with heavy-flavor jets,
a neural network (\rm{NN}) based $b$-tagging
algorithm is applied that
exploits the longer lifetimes of $b$-flavored hadrons in comparison to their lighter counterparts \cite{bid}.
The inputs to the \rm{NN} combine several characteristic quantities of the jet and associated
tracks to provide a continuous output value that tends towards one for $b$ jets and zero for non-$b$ jets. 
The important input variables are the number of reconstructed  
\rm{SV} in the jet, the invariant mass of charged particles associated with the \rm{SV}
($M_{\text{SV}}$), the number of tracks used to reconstruct the \rm{SV},
the two-dimensional decay length significance of the SV in the plane transverse to the beam,
a weighted combination of the tracks transverse \rm{IP} significances,
and the probability that the tracks from the jet originate from the PV, which is referred to as the 
Jet Lifetime Probability (\rm{JLIP}). 
We require at least one of the jets in the event to have a \rm{NN} output greater than 0.5.  
In the case where the leading jet 
is not tagged, we apply the \rm{NN} selection to sub-leading jets. 
A total of 2,200 (1,015) events with at least one $b$ tagged jet candidate are thus selected in the 
$ee$ ($\mu\mu$) channel. The tagging efficiency for $b$ jets and 
the mistagging rate of light jets are parametrized as 
functions of jet $p_{T}$ and $\eta$, and are about 58\% and 2\%, respectively, averaged over the
kinematics of jets considered in this analysis.

\begin{figure}
\begin{center}\includegraphics[%
  width=7cm,
  height=6.5cm,
  trim=0 60 0 0
  ]{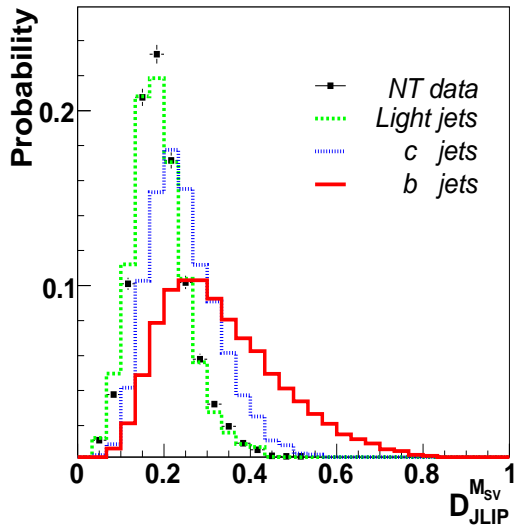}\end{center}

\caption{\label{fig:mjl} The probability densities of the \mjl\ discri-
minant for $b$, $c$ and light jets passing the \rm{NN} $b$ tagging 
requirement. Also shown is the distribution for the neg-
ative tagged (\rm{NT}) jets in data, described in the text.}
\end{figure}

To further separate $b$ jets from $c$ and light jets, 
we construct a discriminant (\mjl) from the combination of $M_{\text{SV}}$ and \rm{JLIP}, 
$\mjl\ \equiv (M_{\text{SV}}/10~\GeV - \ln(\text{JLIP})/40)$.
The relative weights of the variables are selected based on studies of simulated data
to maximize rejection of $c$ and light quark jets.
The mass $M_{\text{SV}}$ provides good discrimination between $b$, $c$, and 
light jets due to the different masses of the quarks. 
Jets from $b$ quarks usually have large values of $-\ln(\text{JLIP})$, 
while light jets mostly have small values, as their tracks originate from the PV.
The average efficiency for the $b$ jets in data to have a well-defined \mjl\ output is about 68\%,
which is due to the finite efficiency for a $b$ jet to have a reconstructed \rm{SV}.
Figure~\ref{fig:mjl} shows the normalized distributions of \mjl\ for jets of different flavors
after the \rm{NN} $b$ tagging requirement. The discriminant
\mjl\ separates well between $b$, $c$, and light jets. 
Figure~\ref{fig:mjl} also shows the \mjl\ distribution of the
tagged jets derived from a light jet enriched data sample, 
referred to as negatively tagged (\rm{NT}) data. 
\rm{NT} jets have negative 
values for some of the inputs for the \rm{NN} algorithm~\cite{bid}
such as decay length significance and \rm{IP} which are caused
by the detector resolution effects. 
We estimate the $b$ jet contamination in the NT data using a maximum
likelihood fit and subtract its contribution.
The template shapes in the corrected NT data and the light jets
in Monte Carlo (MC) simulation look similar and the small difference is taken as
a systematic uncertainty.

The dominant background to $Z+{\rm jet}$ production arises
from multijet (\rm{MJ}) events in which jets are mis-reconstructed
as leptons, especially in the $ee$ channel. 
This instrumental background is estimated from data. 
We use \rm{MJ}-enriched data samples that pass all event selection requirements,
but fail some of the lepton quality criteria, to determine the kinematic shape of the 
background distribution.
For the $\ee$ channel, the \rm{MJ} sample is obtained by inverting the
shower shape requirements and relaxing other identification criteria on the 
electron candidates.
For the $\mumu$ channel, the \rm{MJ} sample consists of events with muon candidates 
that fail the isolation criteria.

Smaller background contributions arise from 
top quark pair ($\ttbar$) and diboson ($WW$, $WZ$, $ZZ$) production,
which contain two leptons in the final state. These backgrounds 
are estimated using MC simulations with the cross sections rescaled to match
theoretical calculations~\cite{bgd1,bgd2}. We simulate 
inclusive diboson production with \pythia~\cite{pythia}.
Events from $Z+{\rm jet}$ and $t\bar{t}$ processes are generated with
\alpgen~\cite{alpgen}, interfaced with \pythia~for initial and final state
radiation and for hadronization. 
For these events, a matching procedure is used to avoid
double counting of partons produced by \alpgen\ and those
subsequently added by the showering in \pythia.
The $Z+$jets samples consist of $Z+$light jets and
a $Z+$ heavy-flavor component, which includes $Z+b\bar{b}(c\bar{c})$ production.
All simulations use the {\sc CTEQ6L1}~\cite{cteq6} 
parton distribution functions (PDFs).
All samples are processed using a
detector simulation based on \geant~\cite{geant} and the same
off\-line reconstruction algorithms as for data. Events from
randomly chosen beam crossings are overlaid on the simulated events to
reproduce the effect of multiple $\ppbar$ interactions and detector noise.
The normalizations of the simulated and the \rm{MJ} backgrounds are
adjusted by scale factors determined from a fit to the $\mll$ distributions
in the inclusive untagged sample.
The background fraction in the $ee$ channel is about 18\% for both the inclusive untagged and tagged samples, and is 
dominated by the \rm{MJ} background. The $\mumu$ channel has a higher purity,
with a background fraction of only about 0.8\% in the untagged and tagged samples.

\begin{figure}
\begin{center}\includegraphics[%
  width=8.5cm,
  height=7cm,
  keepaspectratio,
  trim=0 40 0 0
  ]{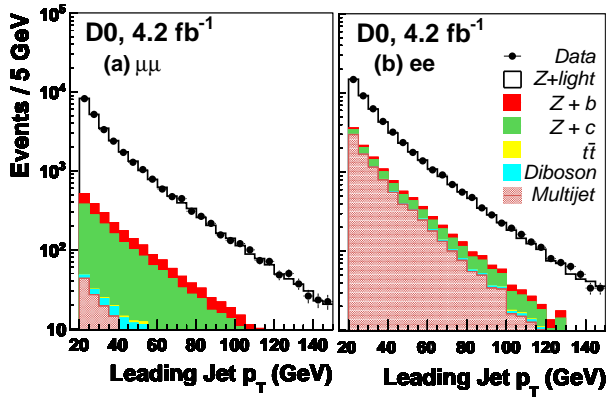}\end{center}  
  
\caption{\label{fig:jetpt} (color online) The observed $p_{T}$ distribution of 
the leading jet in the (a) $\mu\mu$ and (b) $ee$ channel comp-
ared with the SM prediction. The uncertainties on the data points 
are statistical, and the prediction is normal-
ized to the data, as described in the text.}
\end{figure}

Corrections are applied to the simulated events to improve the 
MC modeling. The simulated $Z \rightarrow \mu\mu$ events are weighted
with trigger efficiencies measured in data. For the $\ee$ channel, no
correction is applied as the corresponding trigger is
nearly 100\% efficient. Lepton identification efficiencies are corrected as
a function of $\eta$, azimuthal angle $\phi$, and the 
$z$ position of the PV. Jet energies are smeared
to reproduce the resolution observed in data, and the efficiency for reconstructing a jet
is corrected to match the one in data.
The simulated $Z$ boson events are reweighted
such that the $\pt$ distribution of the $Z$ boson is consistent with the
observed distribution. 
Figures~\ref{fig:jetpt}(a,b) show the $\pt$ distribution of the leading jet in data
compared with the expectation from simulation for $Z+{\rm jets}$ inclusive events and the associated
contributions in each channel. The dominant contribution comes from
$Z+$light jet production.

\begin{figure}
\begin{center}\includegraphics[%
  width=7cm,
  height=6.5cm,
  trim=0 40 0 0
  ]{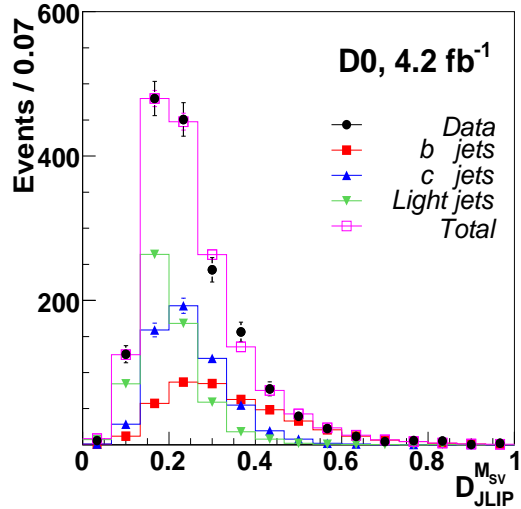}\end{center}  
\caption{\label{fig:fitresult}(color online) The \mjl\ discriminant distribut- 
ion of events in the combined sample. The distributions 
of the $b$, $c$, and light jets are weighted by the fractions found from the fit. 
Uncertainties are statistical only. }
\end{figure}

In order to measure the fraction of events with different jet flavors in the final selected sample,
we perform a binned maximum likelihood fit to the \mjl\ distribution in data using a combination of
the light, $c$, and $b$ flavor jet templates.
Before the fit, we subtract the non-($Z+{\rm jet}$) background contributions.
A total of 970 (630) events remains in the $ee$ ($\mu\mu$) channel passing 
all selection requirements and after the 
background subtraction.
The $b$ and $c$ jet \mjl\ templates are taken from MC simulations
with correction
factors applied to account for the differences in data and MC efficiencies.
The light jet template is obtained from the higher statistics NT data described
earlier. The jet flavor fractions obtained in the $ee$ and $\mu\mu$ channels are 
shown in Table \ref{tab:fractions}, where the 
uncertainties are from the fit due to the data and template statistics.
The relative light and $c$ quark fractions are not tightly constrained by the data. 
The $b$ jet fraction is, however, largely insensitive to variations in the relative amount
of light and $c$ jets.
Since the individual samples yield consistent results, we 
combine the $\ee$ and $\mumu$ samples and remeasure the fractions using an independent fit. 
The \mjl\ distributions in the two data samples used 
for fitting agree after background subtraction.
The last column of Table \ref{tab:fractions} gives the results of the jet flavor fractions 
from the combined sample.
Figure~\ref{fig:fitresult} shows the combined \mjl\ distribution of $b$-tagged jets for data 
along with the fitted contributions from the light (NT data), $c$ and $b$ jets.

The extracted jet flavor fractions are
used to determine the ratio $\sigma(Z+b{\rm\, jet})/\sigma(Z+{\rm jet})$ as follows:\\
\begin{eqnarray}
\frac{\sigma(Z+b{\rm\, jet})}{\sigma(Z+{\rm jet})} =
\frac{N_b}{N_{\text{incl}}\,\epsilon_{b}^{\text{tag}}\,\epsilon_{b/\text{incl}}^{\text{reco}}},
\end{eqnarray}
\noindent where $N_{\text{incl}}$ is the total number of
$Z+{\rm jet}$ events before any tagging requirement, $N_b$ is the number of $Z+b{\rm\, jet}$ events
obtained from the \mjl\ fit, $\epsilon_{b}^{\text{tag}}$ is the overall \mjl\ efficiency for $b$ jets,
which combines the efficiencies for taggability, \rm{NN} tagger and \mjl\ selection, and
$\epsilon_{b/\text{incl}}^{\text{reco}}$ accounts for the difference between
$b$ and inclusive jet reconstruction efficiencies.

\begin{table}
\caption{Jet flavor fractions obtained from template fitting in the dielectron, dimuon and combined channels, along with statistical uncertainties.\label{tab:fractions}}
\begin{center}\begin{tabular}{lccc}
\hline
\hline
Channel& $\mu\mu$ & $ee$ & Combined\tabularnewline
\hline
Events& 630 & 970 & 1600\tabularnewline
$Z+b$ & $0.248\pm0.042$ & $0.267\pm0.036$&$0.259\pm0.028$\tabularnewline
$Z+c$ & $0.253\pm0.073$ & $0.364\pm0.064$&$0.359\pm0.049$\tabularnewline
$Z+\rm{light}$ & $0.499\pm0.058$ & $0.369\pm0.049$ &$0.382\pm0.038$\tabularnewline
\hline
\hline
\end{tabular}\end{center}
\end{table}

Several experimental uncertainties cancel out in the measurement of 
$\sigma(Z+b{\rm\, jet})/\sigma(Z+{\rm jet})$, including the
uncertainties on the luminosity, and trigger, lepton, and some jet identification
efficiencies. 
The two largest remaining sources of systematic uncertainty
are uncertainties in the \mjl\ efficiency and in the shape of the \mjl\ templates
used for the extraction of the $b$ jet fraction. 
Variation in \mjl\ efficiency by one standard deviation results in an 
uncertainty of 3.7\% on the final result.
The uncertainty due to the shape of the templates (4.2\%) is estimated by
using an alternate light jet template from MC, by changing the $b$ quark fragmentation function~\cite{pythia}, and
by varying the fraction of merged heavy quarks ($b\bar{b}$, $c\bar{c}$) inside the jet.
Other important sources of uncertainty are the $b$ tagging efficiency (2.4\%),
the $b$ jet energy scale (2\%), and reconstruction efficiency (3.2\%). 
The total systematic uncertainty on the measurement of the ratio is 7.7\%.
The final result is
\begin{equation}
\frac{\sigma(Z+b{\rm\, jet})}{\sigma(Z+{\rm jet})} = 0.0193\pm0.0022({\rm {stat}})\pm0.0015({\rm {syst})}, \nonumber
\end{equation}
which is consistent with the ratios obtained separately for the
two channels. This measurement is the most precise to date. 
For the kinematic region considered in the analysis,
an NLO \mcfm~\cite{Campbell} prediction for the ratio 
yields $0.0192\pm0.0022$;
this is obtained for the renormalization and factorization
scales $Q^2_R=Q^2_F=m_Z^2$ ($m_Z$ being the $Z$ boson mass),
and with the MSTW2008 PDFs~\cite{mstw}. The prediction decreases 
by 3.6\% when the effects from detector response, resolution 
as well as hadronization and underlying event are taken into account.

In summary, we have performed the most precise measurement to date of the ratio of 
the cross section for $Z$ boson production in association with at least one $b$ jet 
to the inclusive $Z+{\rm jet}$ cross section,
considering final states with $Z \rightarrow ee$ and $Z \rightarrow \mu\mu$ 
and jets with $\pt > 20~\GeV$ and $|\eta| \leq 2.5$.
The combined measurement of the ratio yields 
%$0.0193\pm0.0022$\,(stat)\,$\pm0.0015$\,(syst),
$0.0193\pm0.0027$,
which is consistent with NLO QCD calculations.

%\input acknowledgement_paragraph_r2.tex   % input acknowledgement
% acknowledgement.tex                             24 August 2010
%
We thank the staffs at Fermilab and collaborating institutions,
and acknowledge support from the
DOE and NSF (USA);
CEA and CNRS/IN2P3 (France);
FASI, Rosatom and RFBR (Russia);
CNPq, FAPERJ, FAPESP and FUNDUNESP (Brazil);
DAE and DST (India);
Colciencias (Colombia);
CONACyT (Mexico);
KRF and KOSEF (Korea);
CONICET and UBACyT (Argentina);
FOM (The Netherlands);
STFC and the Royal Society (United Kingdom);
MSMT and GACR (Czech Republic);
CRC Program and NSERC (Canada);
BMBF and DFG (Germany);
SFI (Ireland);
The Swedish Research Council (Sweden);
and
CAS and CNSF (China).
We thank the author of MCFM for the help with useful discussion on the
theoretical calculations.

\end{document}